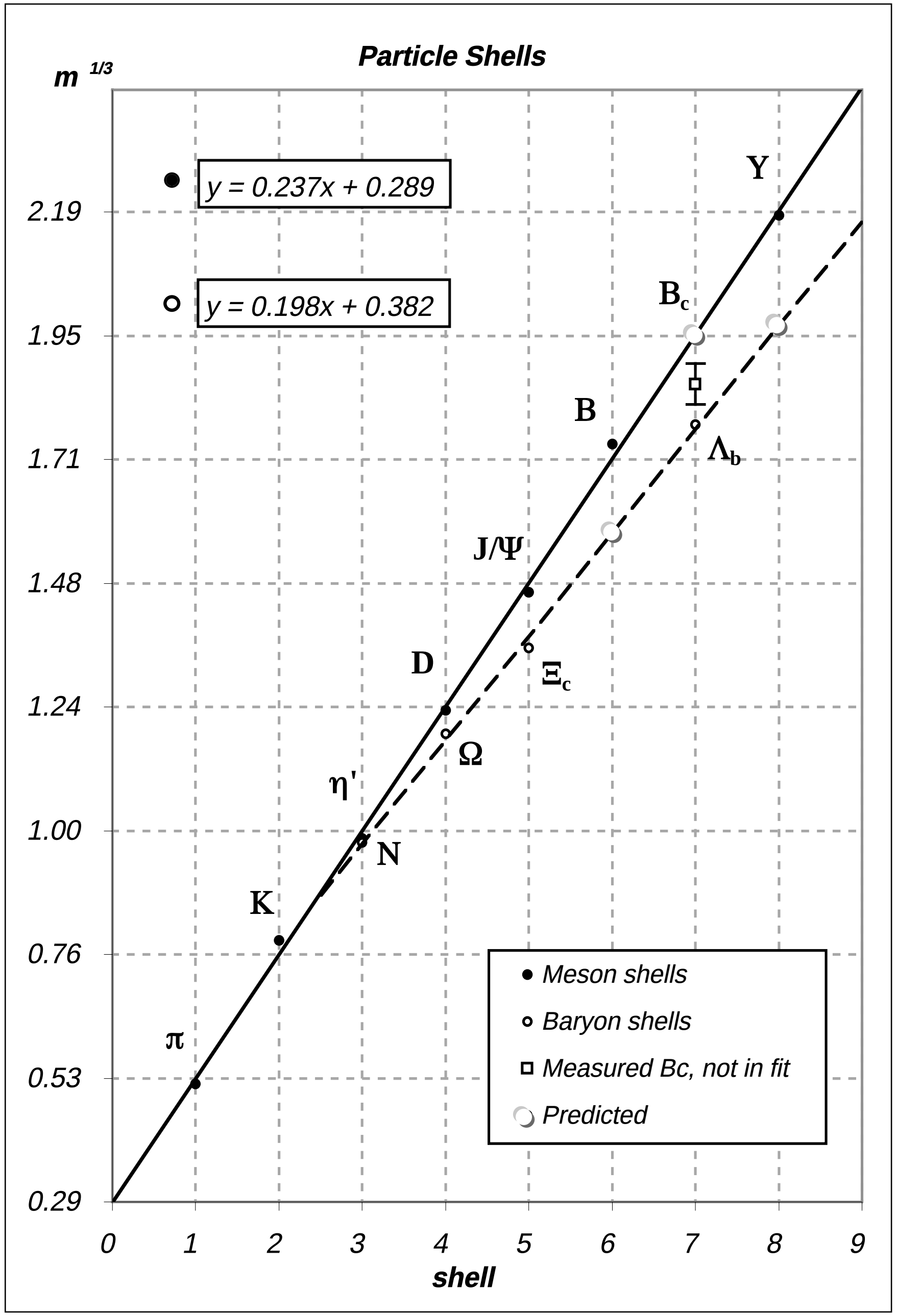

# Particles and Shells

## Paolo Palazzi, CERN

### Abstract

The current understanding of particle masses in terms of quarks and their binding energy is not satisfactory. Both in atoms and in nuclei the organizing principle of stability is the shell structure, while this does not seem to play any role for particles. In order to explore the possibility that shells might also be relevant at this inner level of aggregation, atomic and nuclear stability are expressed by "stablines", alignments of the 1/3 power of the total number of constituents of the most stable configurations. Could similar patterns be found in the particle spectrum? By analyzing the distribution of particle lifetimes as a function of mass, stability peaks are recognized for mesons and for baryons and indeed the cube roots of their masses follow two distinct stablines. Such alignments would be a strong indication that the particles themselves are shell structured assuming only that each constituent contributes a constant amount to the total mass. This is incompatible with the prevalent view that the partons —real physical constituents seen in deep-inelastic scattering experiments—are the quarks. The mass of the $B_c$ predicted by interpolation with the meson stabline is 7.4 ±0.2 GeV. On the baryon stabline two missing states are predicted at 3.9 and 7.6 GeV.

Address correspondence to: paolo.palazzi@cern.ch

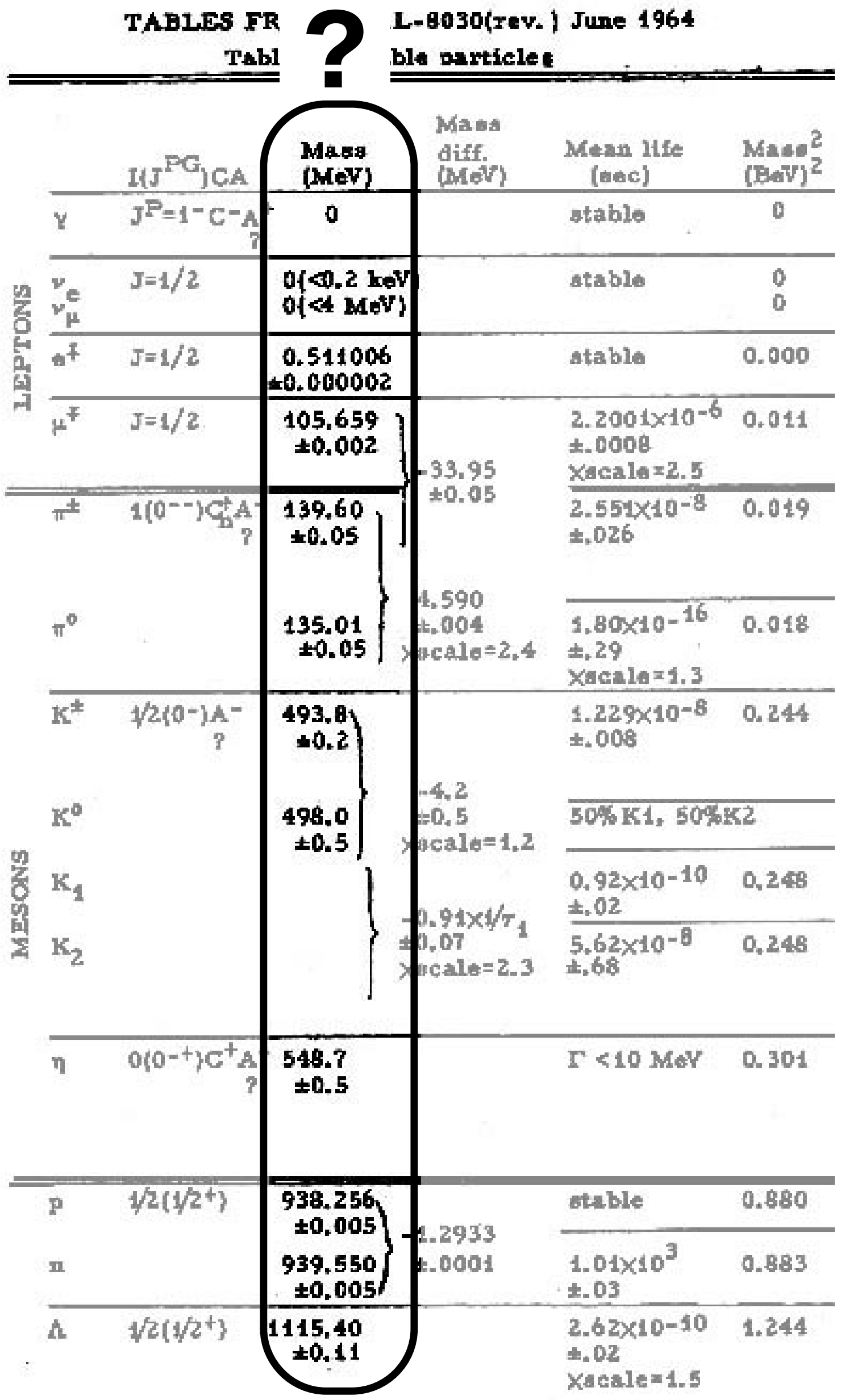

TABLES FR ? L-8030(rev.) June 1964
Tabl ble particles

| | | $I(J^{PG})CA$ | Mass (MeV) | Mass diff. (MeV) | Mean life (sec) | $Mass^2$ $(BeV)^2$ |
|---|---|---|---|---|---|---|
| | γ | $J^P=1^- C^- A$ ? | 0 | | stable | 0 |
| LEPTONS | $\nu_e$ $\nu_\mu$ | J=1/2 | 0(<0.2 keV) 0(<4 MeV) | | stable | 0 0 |
| | $e^\mp$ | J=1/2 | 0.511006 ±0.000002 | | stable | 0.000 |
| | $\mu^\mp$ | J=1/2 | 105.659 ±0.002 | -33.95 ±0.05 | $2.2001\times10^{-6}$ ±.0008 ×scale=2.5 | 0.011 |
| MESONS | $\pi^\pm$ | $1(0^{--})C_n^+A$ ? | 139.60 ±0.05 | | $2.551\times10^{-8}$ ±.026 | 0.019 |
| | $\pi^0$ | | 135.01 ±0.05 | 4.590 ±.004 ×scale=2.4 | $1.80\times10^{-16}$ ±.29 ×scale=1.3 | 0.018 |
| | $K^\pm$ | $1/2(0^-)A^-$ ? | 493.8 ±0.2 | | $1.229\times10^{-8}$ ±.008 | 0.244 |
| | $K^0$ | | 498.0 ±0.5 | -4.2 ±0.5 ×scale=1.2 | 50% K1, 50%K2 | |
| | $K_1$ | | | | $0.92\times10^{-10}$ ±.02 | 0.248 |
| | $K_2$ | | | $-0.91\times1/\tau_1$ ±0.07 ×scale=2.3 | $5.62\times10^{-8}$ ±.68 | 0.248 |
| | η | $0(0^{-+})C^+A$ ? | 548.7 ±0.5 | | Γ <10 MeV | 0.301 |
| | p | $1/2(1/2^+)$ | 938.256 ±0.005 | -1.2933 ±.0001 | stable | 0.880 |
| | n | | 939.550 ±0.005 | | $1.01\times10^3$ ±.03 | 0.883 |
| | Λ | $1/2(1/2^+)$ | 1115.40 ±0.11 | | $2.62\times10^{-10}$ ±.02 ×scale=1.5 | 1.244 |

**Fig. 1.** The mystery of particle masses.

## 1. The Mystery of Particle Masses

Among the mysteries of particle physics, one of the most elusive is the mass spectrum. More than 50 years ago, Y. Nambu proposed an empirical rule with a mass unit close to one quarter of the mass of the pion, and noticed among other things that the muon mass was 3/4 of the pion mass, and that the boson and fermion masses were respectively even and odd multiples of this mass unit [1]. E. Jensen formulated the same idea in 1980 [2] treating the masses of many more particles with a thorough statistical analysis of the residuals, but without providing any physical interpretation.

The quark model has been extremely successful in accounting for a number of properties of the particle spectrum, for interactions and decays, but not for masses. The masses attributed to the quarks are questionable, and so is the evaluation of binding energies required to obtain the masses of particles. The Zweig rule, the Cabibbo angle and the absence of free quarks may seem good reasons to conclude that quarks represent collective valence properties of some kind, but are not real physical objects. The results of deep-inelastic scattering experiments are compatible with the charges and other properties of the quarks, and for that reason "partons = quarks" is currently the conventional view, but the inability to explain particle masses suggests that alternatives may be worth exploring.

"Can 35 pionic mass intervals among related resonances be accidental?" titled M. H. Mac Gregor [3] in 1980. In "An elementary particle constituent-quark model" [4] published in 1989, the same author plotted the masses of mesons and baryons, and showed that "particle masses occur in … mass bands" featuring "a mixture of spins, parities, isotopic spin … all occurring in a single mass band". "The pattern is not a rotational-energy type of systematic, and can be explained with clusters of spin 1/2 quarks occurring in various spin-up spin-down configurations".

In light of the known shells of atomic and nuclear structure theory, it is of interest to ask whether or not the discrete mass differences and bands mentioned above might also indicate the existence of shells at the particle level. Related ideas combined with the quark model have been published by J. W. Moffat in 1976 [5], while the present study is a model-independent search for shell signatures across the particle spectrum.

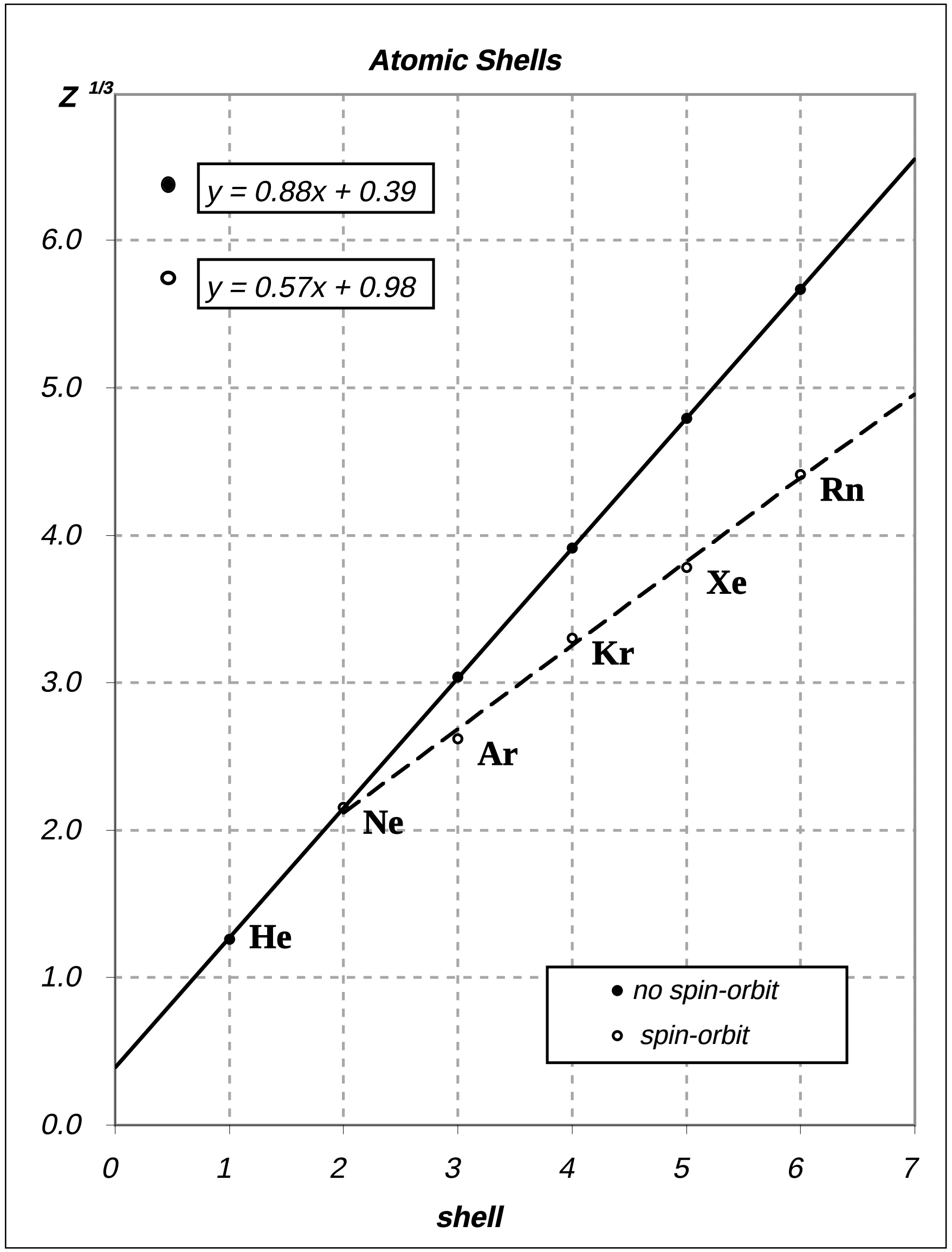

**Fig. 2.** Atomic stabline, the cube root of the total number of electrons for closed shell configurations, without and with spin-orbit coupling, and line fits to each series.

## 2. Atomic Shells

The Atomic Table of the Elements of D. I. Mendeleyev began as a set of cards, with the elements and atomic weights on their faces, which he would shuffle around as if in a game. In this way he uncovered the property of periodicity—namely that the elements, when arranged in order of their atomic weights, have such similar properties when aligned in certain regular and recurring intervals, that they may be grouped into families.

A complete understanding of this periodicity was made possible by the solution of the Schrödinger equation for the hydrogen atom, where three quantum numbers arise from the 3D space geometry of the wave function, and a fourth one from the spin of the electron. The quantum numbers set limits on the number of electrons that can occupy a given state and therefore give insight into the build-up of the periodic table of the elements.

| | |
|---|---|
| For each value of the principal number | $N = 1, 2, 3,..$ |
| the orbital number can assume N values | $l = 0, 1, 2, .. ,N-1$ |
| the magnetic number 2l+1 values | $m_l = -l, -l+1,..0.., l-1, l$ |
| and the spin number 2 values | $s = -1/2, 1/2$ |

Each value of the principal number N defines a shell containing a maximum of $2N^2$ electrons, the series:

2, 8, 18, 32, 50, 72, 98,..

The total number of electrons up to shell N is the integral series 2N(N+1)(2N+1)/6, i.e.:

Z = 2, 10, 28, 60, 110, 182, 280,..

The cube root of this series as a function of N is a straight line with a small intercept. The experimental values in column VIII of the periodic table, corresponding to the atomic numbers of the noble gases, are actually lower starting from the third shell because of spin-orbit interactions, and they line up on another stabline with a lower slope.

*The stabline, an alignment of the cube root of the total number of constituent, corresponds to the most stable configurations of electrons in each row of the periodic table. While more than 100 other configurations of electrons are known, it is these 6 inert gases that are the main organizing principle of atomic stability.*

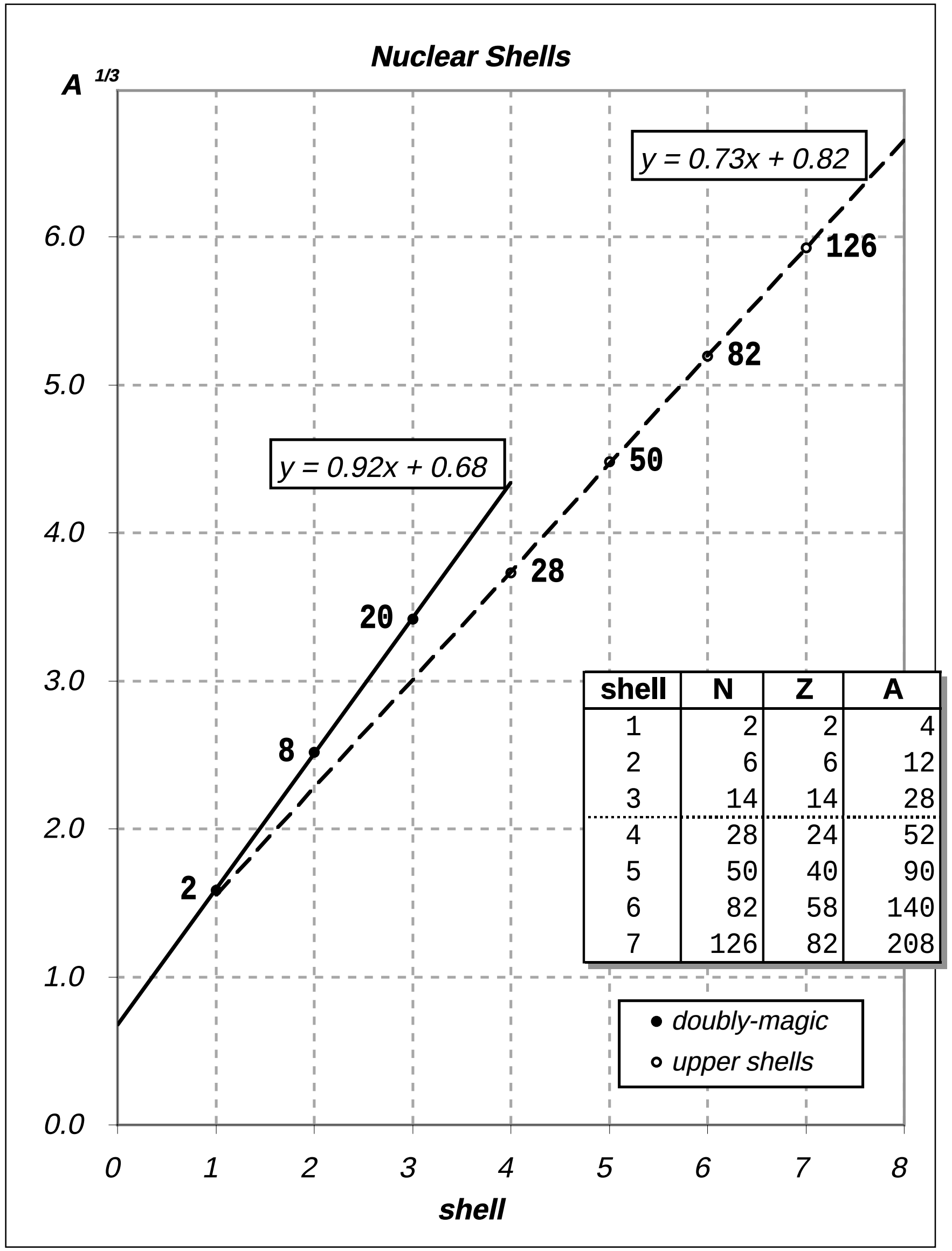

| shell | N | Z | A |
|---|---|---|---|
| 1 | 2 | 2 | 4 |
| 2 | 6 | 6 | 12 |
| 3 | 14 | 14 | 28 |
| 4 | 28 | 24 | 52 |
| 5 | 50 | 40 | 90 |
| 6 | 82 | 58 | 140 |
| 7 | 126 | 82 | 208 |

**Fig. 3.** Nuclear stablines, fitting the cube root of A. The black points refer to the doubly-magic shells 1, 2 and 3, the white points are shells 4 and above. The points are tagged with the corresponding neutron magic number.

## 3. Nuclear Shells

In 1948, M. Göppert-Mayer, working on the origin of the elements, was improving on the findings by W. M. Elsasser, who had already noticed in 1933 that the most abundant elements had particular numbers of neutrons or protons in their nuclei. Göppert-Mayer had more data available and found stronger and more diversified evidence. These “magic numbers” suggested the idea of stable shells in nuclei similar to shells in atoms, but the prevailing wisdom was that a shell structure in nuclei was unlikely, due to the short range of nuclear forces compared to the long-range Coulomb forces.

A further difficulty was that the magic numbers did not fit simple-minded quantum mechanical ideas of shell structure. While she was struggling to fit 28 as the fourth magic number, Fermi helped her by asking the key question, "Is there any indication of spin-orbit coupling?" and thus was born the spin-orbit coupling shell model of nuclei, the ancestor of the Independent-Particle (IP) model, currently the dominant theoretical paradigm in nuclear physics.

Spin-orbit coupling is not the only complication of nuclear stability. In nuclei there are two kinds of constituent, and the same magic number series identifies relative stability peaks for both protons and neutrons, although the effect is more pronounced with neutrons. Up to the third magic number the most stable nuclei are doubly magic, with an equal number of neutrons and protons, while further up the stability is maximal for a growing neutron excess.

The plot of the cube root of the mass number A of the most stable nuclei corresponding to the neutron magic number series gives a feeling for the problem with magic number 28. Up to the third shell the points line up precisely, further up maximal stability shifts to a different line that meets the lower one at the first shell. (An exhaustive treatment of nuclear stability from this viewpoint is beyond the scope of this paper, and will be developed separately).

*As is the case for atomic structure, two distinct stablines reflect the nature of nuclear stability. The first one fits the doubly-magic nuclei of shells 1, 2 and 3 and is a pure expression of the nuclear Schrodinger equation—without the additional complications of configuration-mixing and the influences of protons on neutron build-up (and vice versa). The second stabline is that which reflects the stability of the larger nuclei, where these complications play a significant role in altering the pure QM sequence.*

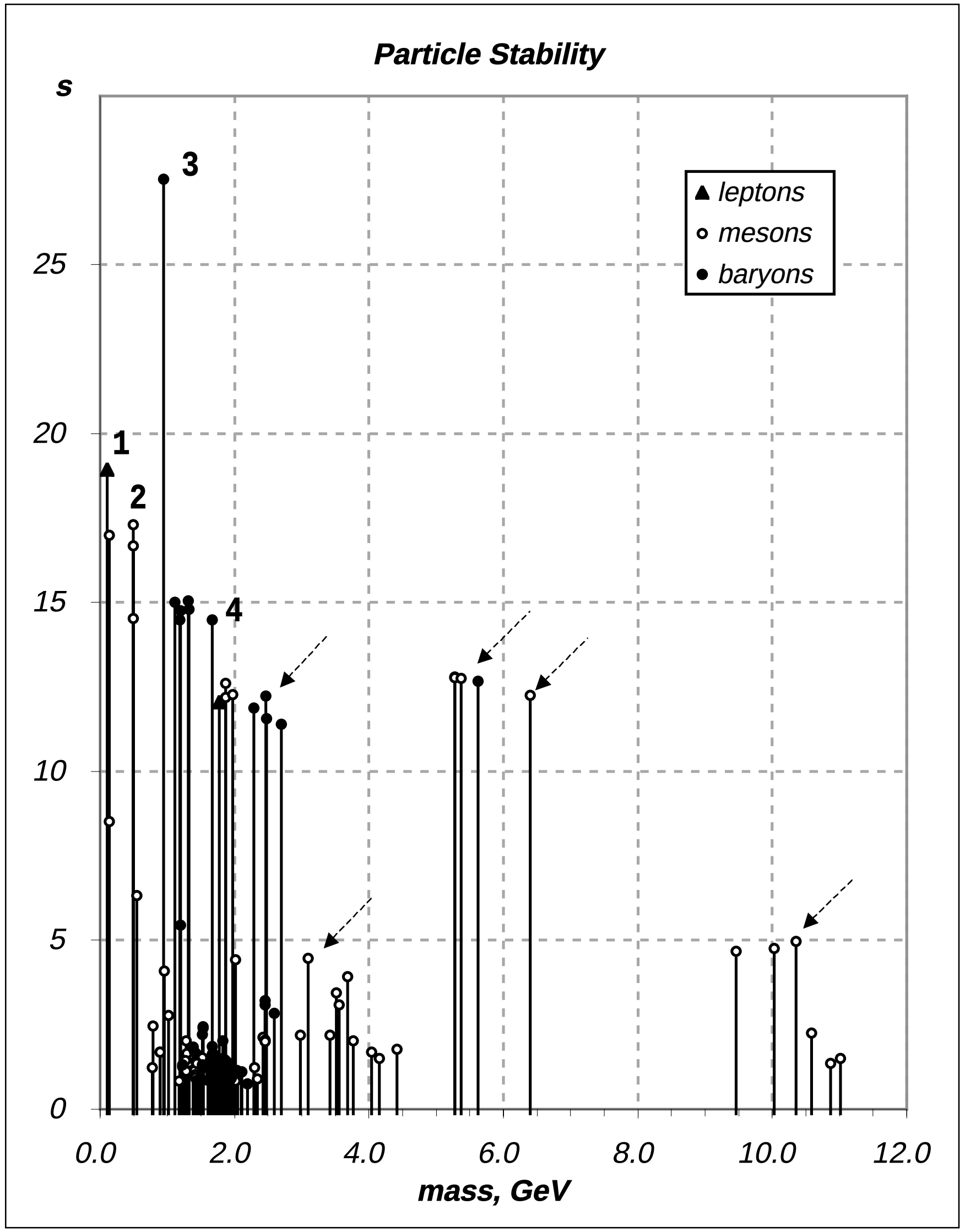

**Fig. 4.** Stability s(i) = $\log_{10}(\lambda(i)/\lambda(Z^o))$ versus the mass for all the particles in the 2002 PDG computer file [6] for which the width is defined, omitting the stable leptons, the $Z^o$ and the W, for a total of 2 leptons, 76 mesons and 61 baryons. Four stability peaks are identified in the low-mass region, corresponding to the $\pi$, K, N and $\Omega$, and more are visible at higher masses, marked with arrows. The $\tau$ lepton is close to a cluster of mesons in peak 4 at stability = 12.

## 4. Particle Stability

It has just been shown that atomic and nuclear stability can be expressed with a pattern based on the volume of concentric spheres with radii increasing by a fixed amount. In both cases spin-orbit coupling distorts the picture, and historically a stability pattern had been found long before it could be explained by a model.

For particles the case is quite different: a model has been around for decades, it is not fully satisfactory, and it does not encourage thinking in terms of shells. Therefore looking for shells with a model-independent analysis seems more appropriate. Starting from the unorthodox assumption that *the number of constituents for each particle is unknown but proportional to the mass*, an analysis of stability versus the mass can be attempted.

Atoms and nuclei are long-lived objects, and for each atomic number many stable isotopes have been studied, producing empirical evidence for the magic numbers. On the contrary, particles are loose structures, instability is the norm and, apart from the proton, only some leptons and the photon are stable. The mass spectrum is sparse, and looking for peaks is a different matter.

The logarithm of the lifetime normalized to the lifetime of the $Z^o$ will be taken as an indicator of particle stability:

$$\mathbf{s(i) = \log_{10}(\lambda(i)/\lambda(Z^o))}$$

Figure 4 is the plot of s(i) against the mass, where i is any particle in the 2002 PDG computer file [6] for which the width has been defined, excluding all stable particles and also the $Z^0$ and the W. Starting from the low masses and ignoring the stable leptons, the first peak just above 100 MeV corresponds to the $\mu$, the $\pi^\pm$ and the $\pi^o$. The second peak is found at about 500 MeV, with the $K^\pm$, the $K^o$ and the $\eta$. The nucleon sticks out as the third peak, around 1 GeV and close to the $\eta$'. After several hyperons and a gap with short-lived resonances, at about 1.7 GeV is peak 4 with the $\Omega$, not far from the D and with the $\tau$ lepton nearby. The $\mu$ and the $\tau$ are on stability peaks 1 and 4, as if the mass generation mechanism were the same for baryons, meson and unstable leptons.

Further stability zones corresponding to the $\Xi_c$ at 2.4 GeV, the $J/\Psi$ at 3.1, the B at 5.3, the $B_c$ at 6.4 and the $\Upsilon$ above 10 are clearly visible. This unconventional depiction of particle masses in relation to lifetimes already reveals certain regularities in terms of peaks and bands. The masses of the first 4 peaks can now be used to probe the presence of stablines, as are evident for atomic and nuclear structure.

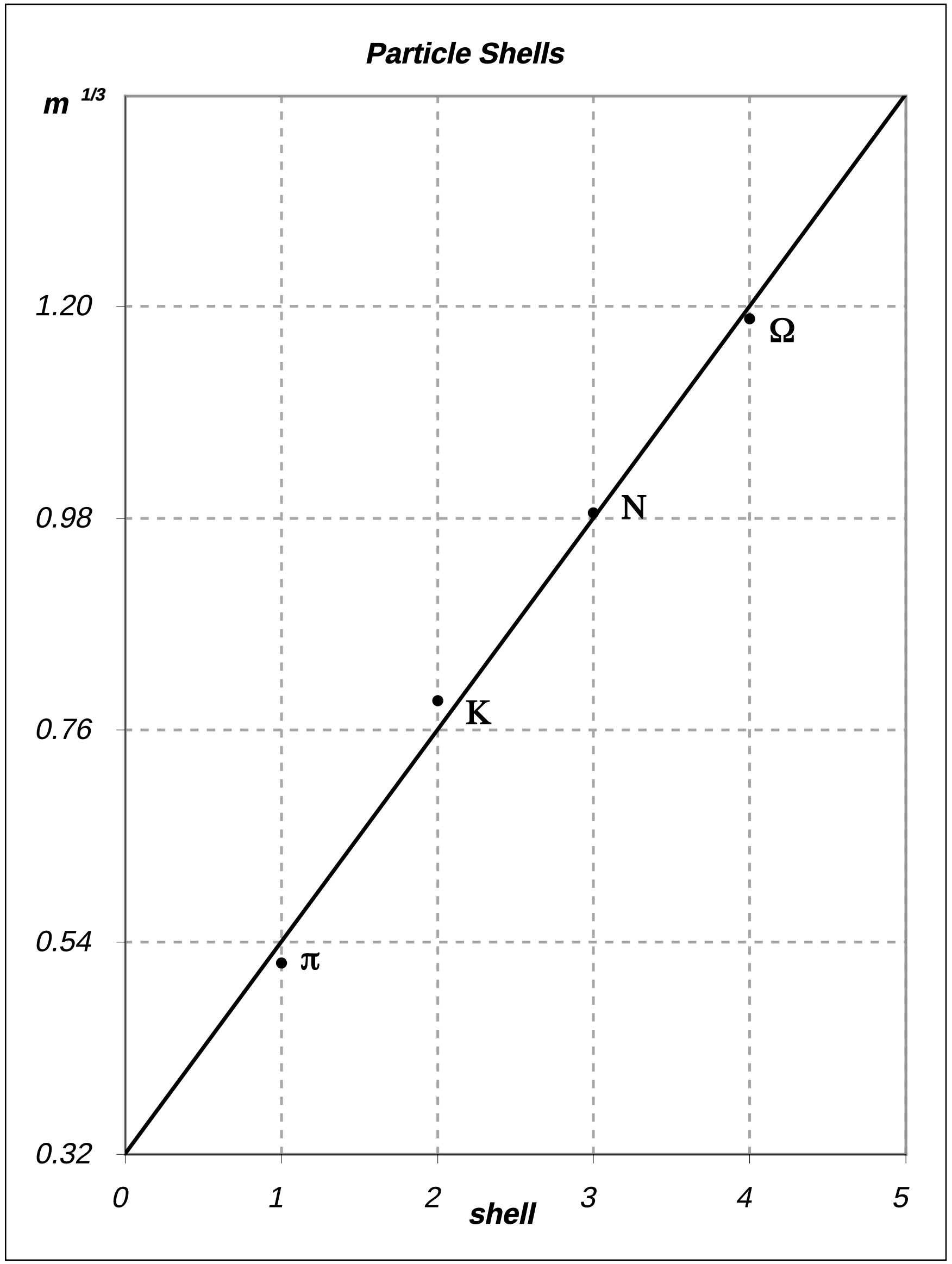

**Fig. 5.** Mass stabline plot of the particles at the four stability peaks identified in figure 4, i.e. π, K, N and Ω. The cube root of the mass in GeV is plotted against the peak serial number. Ad hoc vertical scale adjusted to the shells.

## 5. Particle Shells

Leaving leptons aside for the time being, it seems appropriate to choose as representatives of the four peaks the π, K, N and Ω. A straight line fits the cube root of their masses pretty well, with small but not negligible residuals for the mesons. This may indicate that meson and baryon stability follow distinct but close lines. If so, separate scans for mesons and baryons may reveal a sharper pattern.

### 5.1 Meson Shells

Figure 6 on the next page is the stability plot restricted to the mesons. Eight peaks stick out very clearly, corresponding to the series:

$$( \pi , K , \eta' , D , J/\psi , B , B_c , Y )$$

The stabline fit shown in figure 7 is remarkably good, and the residuals for the π and the K are smaller compared to the fit of figure 5. The measured mass of the $B_c$ is 6.4 ±0.39 ±0.13 GeV, the PDG computer file quotes a combined error of ±0.4 GeV and the point is omitted from the fit. A $B_c$ meson in shell 7 is consistent with the progression of the quark composition across the meson shells, and from the meson stabline the mass of the $B_c$ can be predicted at 7.41 ±0.19 GeV.

### 5.2 Baryon Shells

The baryon stability plot is shown in figure 8: there are no baryons with mass lower than the nucleon, and the plot is sparser. The next two peaks are clearly visible, one corresponding to the isolated Ω and the next one with the $\Xi_c$ surrounded by the $\Lambda_c$ and the $\Omega_c$, then there is a large gap and further up the isolated $\Lambda_b$ . By comparing this series with the meson series in the previous section and aligning the masses, a likely assignment for 8 baryon shells could be:

$$( - , - , N , \Omega , \Xi_c , * , \Lambda_b , * )$$

where the minus sign stands for 'probably nothing' and the asterisk for 'hopefully something'. Figure 9 shows the baryon mass stabline fit. The slope is 15% lower compared to the meson line, which explains the residuals of the fit at figure 5 where the particles chosen for the first attempt of a stabline plot were two mesons and two baryons. No baryon is reported by the PDG at shells 6 and 8, and from the baryon stabline the two missing states can be predicted at 3.88 ±0.1 GeV and 7.63 ±0.1 GeV respectively.

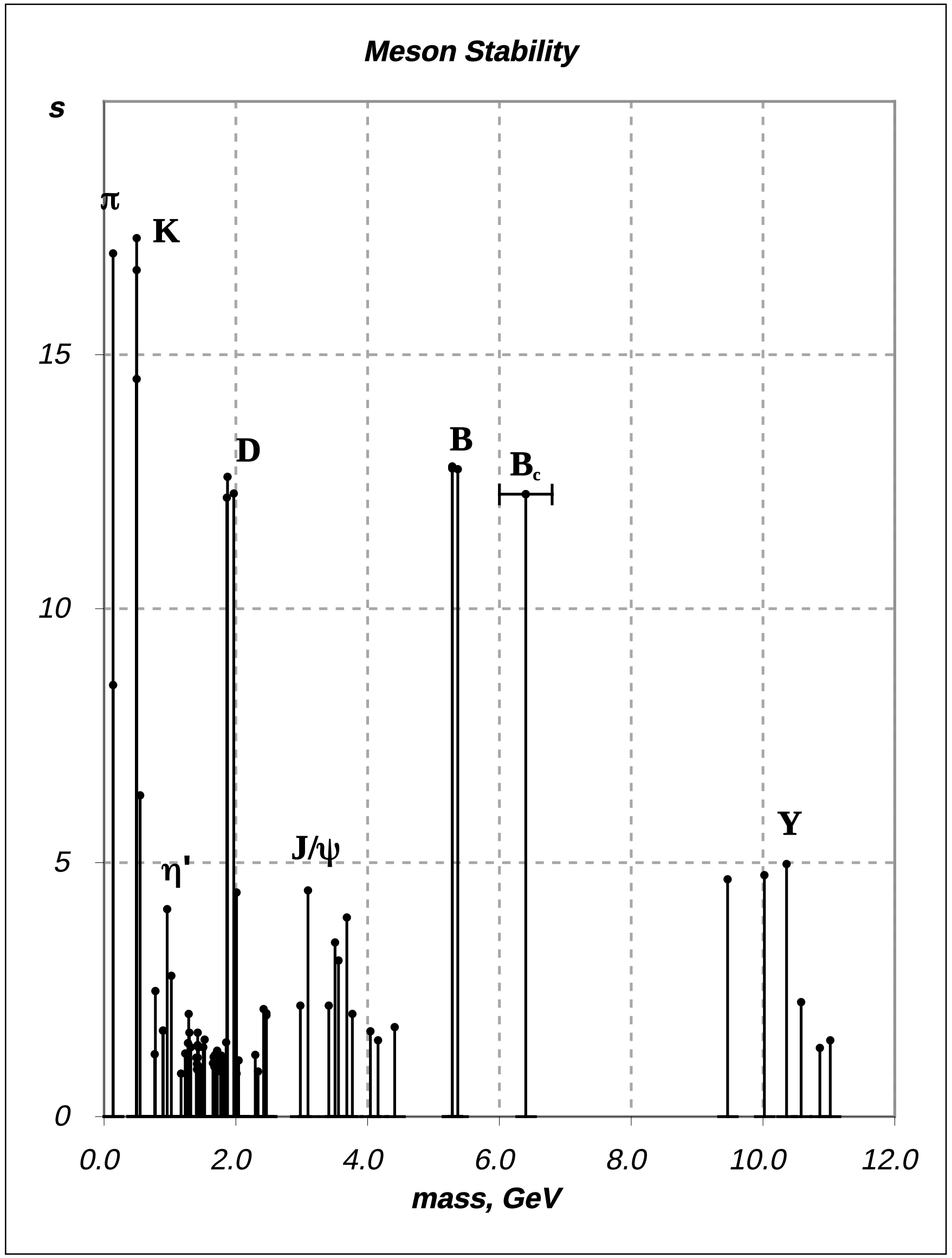

**Fig. 6.** Stability $s(i) = \log_{10}(\lambda(i)/\lambda(Z^o))$ versus the mass for all the mesons in the 2002 PDG computer file. The 8 stability peaks correspond to the masses of the π, K, η', D, ψ, B, $B_c$ and Y. The mass of the $B_c$ has an error of 0.40 GeV.

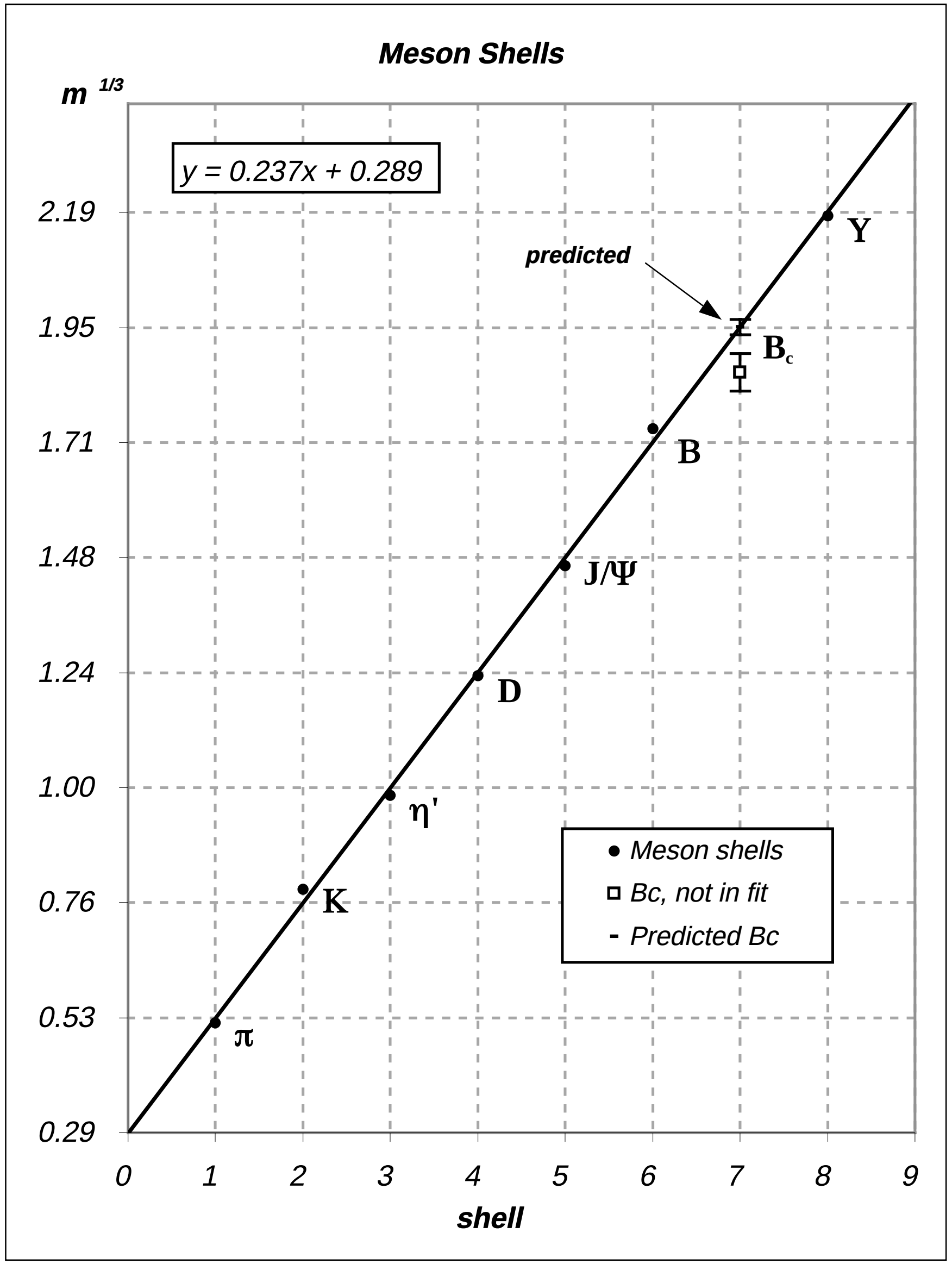

**Fig. 7.** Mass stabline plot of the mesons at the top of the 8 stability peaks of figure 6. The point in shell 7 is not used in the fit. The $B_c$ mass predicted with the stabline is more precise than the measured value. Ad hoc vertical scale adjusted to the meson shells.

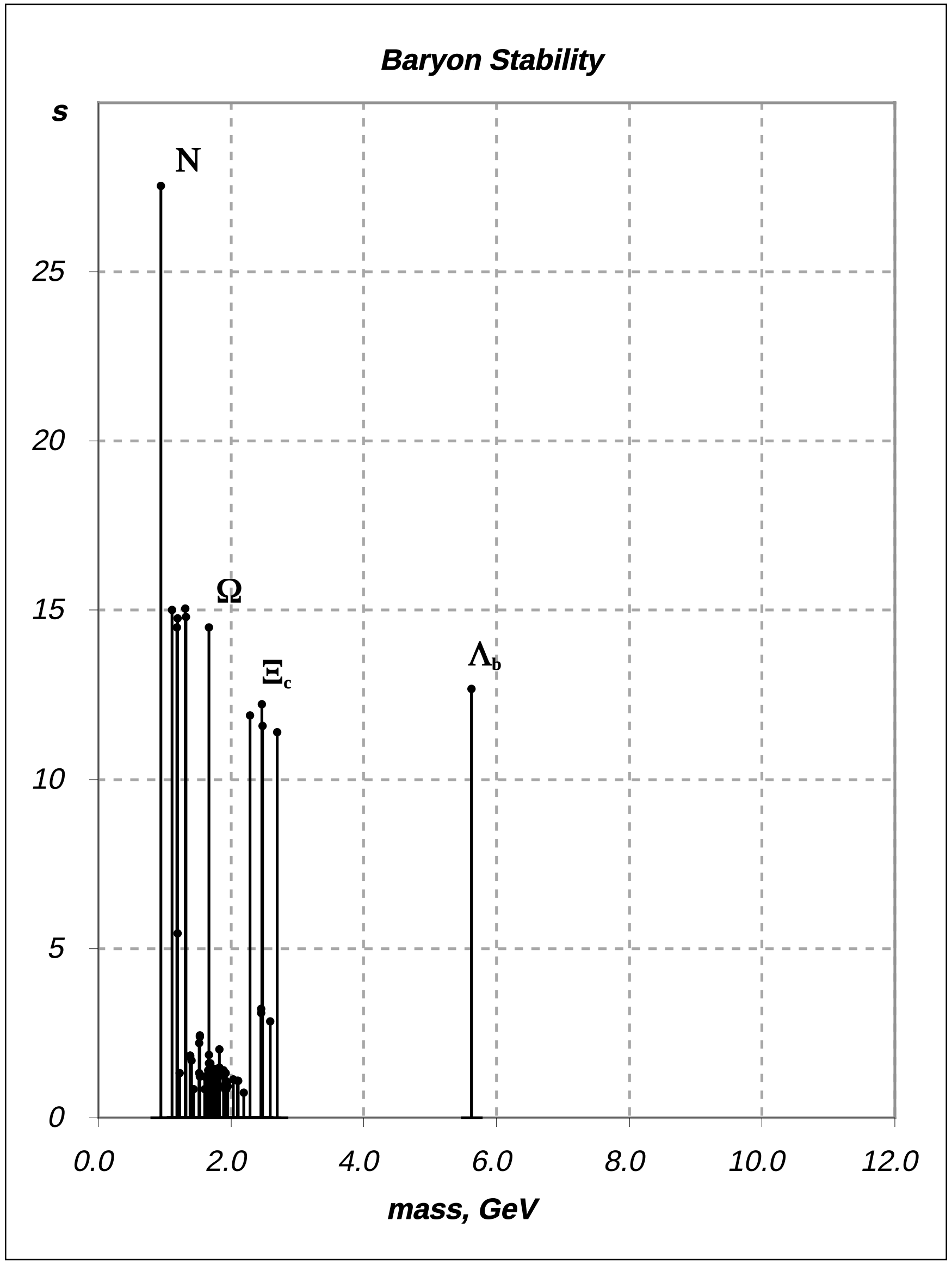

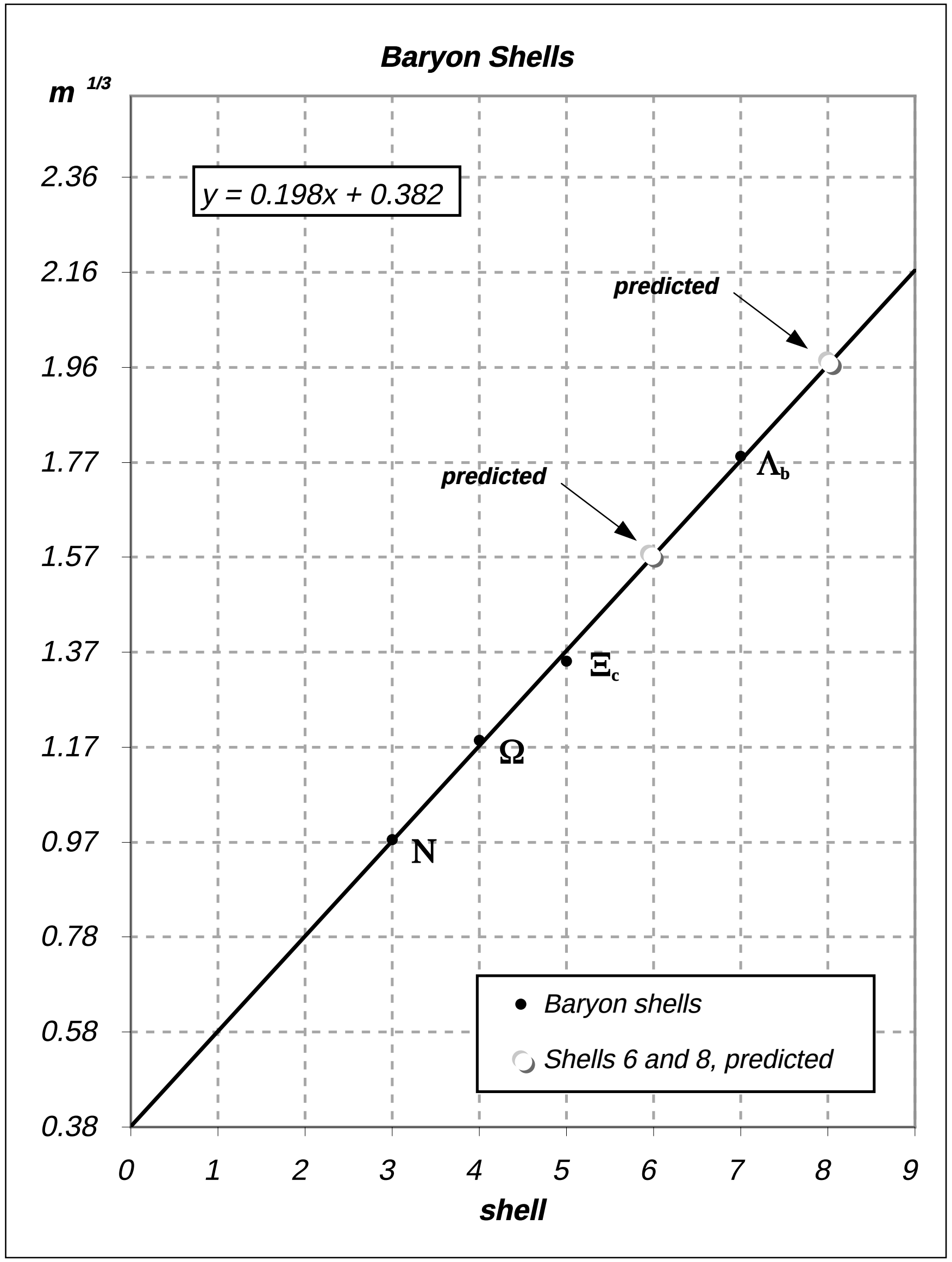

**Fig. 8.** Stability $s(i) = \log_{10}(\lambda(i)/\lambda(Z^o))$ versus the mass for all the baryons in the 2002 PDG computer file. Four stability peaks are visible, at the masses of the N, Ω, $\Xi_c$ and $\Lambda_b$.

**Fig. 9.** Mass stabline plot showing the cube root of the mass of the baryons on the 4 stability peaks of figure 8, with the peak serial number aligned with the mesons. New baryons states are predicted at shells 6 and 8. Ad hoc vertical scale adjusted to the baryon shells.

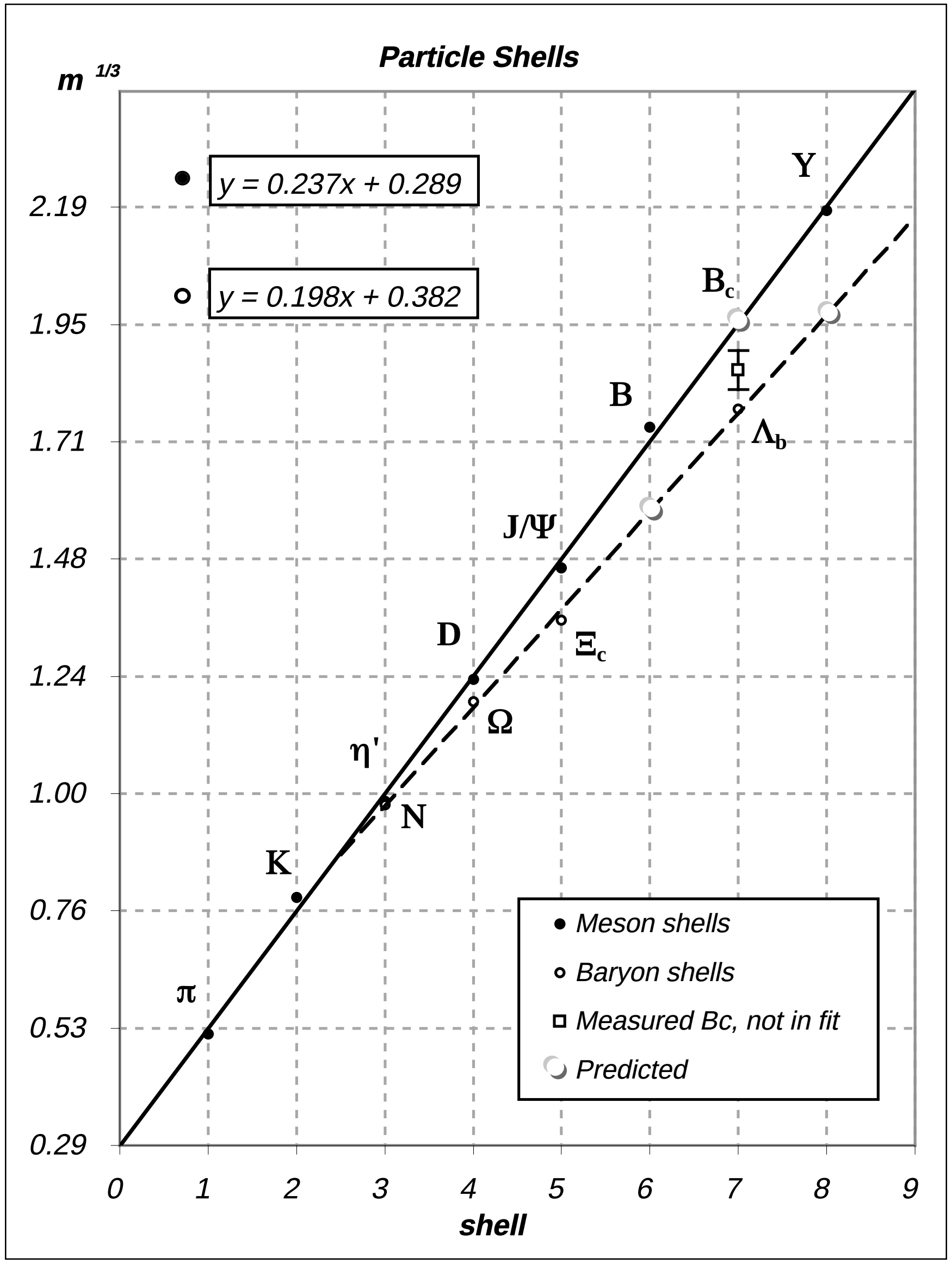

**Fig. 10.** Meson and baryon stablines; the measured $B_c$ is excluded from the meson fit; the $B_c$ mass is predicted with the meson stabline; baryon states at shells 6 and 8 are predicted with the baryon stabline. The η' and N points overlap. Vertical scale adjusted to the meson shells.

## 6. Conclusions

The full evidence for particle shells is presented in figure 10. The pattern seen here indicates a regularity of particle masses that is not immediately obvious from the raw data. Assuming only that the constituents within these particles contribute a constant amount to the total mass, the two stablines strongly suggests a "shell structure" similar to those known at the atomic and nuclear levels: higher masses are obtained by adding more partons, not by replacing quarks with heavier ones. This unconventional approach is incompatible with the prevalent view that identifies the partons—real physical constituents of particles seen in deep-inelastic experiments—with the quarks. Keeping this problem in mind, the next question concerning the precise configuration of such shells can be addressed on the basis of the following observations:

- particles consist of a variable number of constituents arranged in shells and every constituent contributes to the mass by a fixed amount on average;
- the coupling is with anti-parallel spins such that low-spin bound states can be constructed combining a large number of spin 1/2 partons;
- the different slopes of the two stablines suggest that mesons and baryons differ not only for the even or odd number of partons, but also for the spatial arrangement, and this might be at the origin of baryon number conservation;
- the lightest baryon, the proton, is already a complex bound state, a specially stable configuration preserved by all baryon decays;

To move on from this list of features to a convincing particle shell model requires postulating what the constituents are and how they interact, computing the masses precisely, including isospin multiplet differences, and understanding decays and lifetimes. The model must also be compatible with the well established aspects of quark systematics, and reproduce flavors as collective properties of bound states and not of the constituents. At the same time, it must account for the distributions and sum rules of deep inelastic scattering. These questions will be addressed in future work.

An alternative explanation could be that the quarks as constituents are just fine, and that these regularities look like shells but are not, being related to some other effect. If so, which effect?

## Data and Analysis

Particle properties have been obtained from various releases of the Review of Particle Properties compiled by the Particle Data Group. The first generation plots were produced by hand, later the data have been entered manually in various forms of computer files, then converted semi-automatically from the PDG data into an ADAMO [7] file and later YaPPI [8]. Various tools have been used for the analysis: APL, FORTRAN programs with HBOOK [9] and ADAMO, and in the end the PDG computer files [6] have been converted to MS Excel. For this analysis error bars are almost always irrelevant. Apart from the case of the $B_c$ , the masses are known with high precision and the number of constituents is an integer, therefore plotting a few points and fitting a line by hand with a ruler would be sufficient. If the pattern is not there there are no shells, if it is present any deviation is physics and not a statistical error.

## Acknowledgements

The author is grateful to W. K. H. Panofsky and R. E. Taylor for hospitality in the stimulating environment of group A at SLAC in 1973 when this analysis was started, revealing 4 shells.

An early version of this work [10] with 5 shells, including also a hypothesis on the nature of the constituents, has been submitted in 1975 to support the application as INFN associate level R5. It was met with considerable skepticism by the members of the jury: "contains unverified hypotheses, is vague, and the model proposed is not suitable to explain fundamental physics facts". This is still true today. In the meantime the author has been vaccinated against the young Mayan astronomer syndrome [11], and a few more particles have been added to the plots.

Discussions with the late A. O. Barut on lepton bound states, with R. Brun and N. D. Cook on nuclear models, and with S. Giani on particle mass systematics have proven to be very useful. N. D. Cook, D. Dallman, S. Giani, M. H. Mac Gregor and F. Palazzi reviewed the draft and proposed various improvements. E. Remiddi suggested to add some predictions.

G. Lacommare and M. Marino have designed an ADAMO database for PDG data including decays, and O. Di Rosa has loaded it with data from the PDG data base. Thanks to G. H. Moorhead for help with APL, C. Cerrina Feroni for Excel and PowerPoint, M. Ruggier for language and typography, D. Birker and J. Vigen for electronic publishing and O. Reusa for MacOSX support.

## Publication

## Revision Record

24-JAN-2003 preprint, original electronic submission;

20-FEB-2003 Revision 1: improved graphics, better PDF generation;

19-MAR-2004 Revision 2: N.D. Cook added as reviewer, "stabline" term introduced, $B_c$ removed from meson stabline fit, $B_c$ mass and two baryon states predicted, graphs reformatted, abstract reorganized.

Laptop-friendly layout formatted with PowerPoint and embedded Excel plots on an Apple Macintosh PowerBook running MacOSX.